\documentclass{statsoc}
\usepackage{geometry}
\usepackage{amsmath}
\usepackage{array}
\usepackage{graphicx}
\usepackage{url} 
\usepackage[latin1]{inputenc}
\usepackage[english]{babel}
\usepackage{graphicx}
\usepackage{enumerate}
\usepackage{booktabs}
\usepackage[linesnumbered,ruled]{algorithm2e}
\usepackage{natbib}

\usepackage{amsmath,amssymb}
\usepackage{natbib}

\newtheorem{Theo}{Theorem}
\newtheorem{Def}{Definition}
\newtheorem{prop}[Def]{Proposition}

\providecommand{\keywords}{\textbf{Keywords: }}

\DeclareMathOperator*{\argminA}{arg\,min} 

\title{A novel wave decomposition for oscillatory signals}

\author{Cristina Rueda, Alejandro Rodr\'iguez-Collado, Yolanda Larriba}
\address{Department of Statistics and Operations Research, Universidad de Valladolid, Valladolid, Spain.}
\email{cristina.rueda@uva.es}

\begin{document}

\begin{abstract}
	Oscillatory systems arise in the different science fields. Complex mathematical formulations with differential
	equations have been proposed to model the dynamics of these systems. While they have the advantage of having a direct physiological meaning, they are not useful in practice as a result of the
	parameter adjustment complexity and the presence of noise. In this paper, 
	a novel decomposition approach in AM-FM components, competing with Fourier and other decompositions is presented. Several interesting theoretical properties are derived including the Ordinary Differential Equations describing the signal. Furthermore, the usefulness in real practice is demonstrate to analyse signals associated to neuron synapses and by addressing other questions in Neuroscience.
\end{abstract}


\section{Introduction}

A system in which a particle or set of particles moves returning to its initial state after a certain period is an oscillatory system and an oscillation is the repetitive variation of a signal, or a measure, associated with the system.  Oscillations occur in physical and biological systems but also in human society. Examples of oscillations include the swinging pendulum, the periodic firing of a nerve, the expression of circadian clock genes, the beating of the human heart, signals in the radio frequency, or business cycles in economics. The periodic motion, characteristic of oscillations, is encountered in all areas of science, and a huge number of investigators, from different disciplines have contributed to the advancement of the field using particular perspectives, as they are aware of diverse real problems specific to their areas. 
The terminology, the fundamental concepts, the principles, the conventions, the methods, and the theory of these perspectives are often quite different.

On the one hand, there is the focus of the signal analyst which emphasizes the time-frequency approach and the development of algorithms to process and analyse observable signals; most researchers in the communication field follow this approach; a recommended reference is the book  \cite{Boa16}. Alternatively, from a more physical focus, a dynamic system is described primarily by a set of differential equations. Basic references are, \cite{Wig15}, \cite{Ash16}, \cite{Pik15}. This approach has been the preferred one by researchers in electrophysiology and Neuroscience.

The statistical perspective is a third approach to the subject, suitable when real and noisy signals are observed. It considers models that assume noise terms and is useful to identify the real signal features. This approach has been preferably adopted in Chronobiology as seen in \cite{Lar19}. \\
\\

Extracting features from an observed signal is the first step towards data analysis and  an efficient algorithm to extract the desired features from the recorded signal is also needed. In the case of oscillatory signals, the main features are the number of oscillatory components and the amplitude or peak time of each oscillation. For instances, for physiological signals, it is well known that signal oscillations contain plenty of information about a person's health condition.  In general, inferring the dynamical information from a time series is challenging. Fourier Decomposition (FD) is a traditional approach to the analysis of such signals; however, if the signal is not composed of harmonic functions, then the approach is not useful to extract the features. For example, respiratory flow signals do not usually oscillate like a sinusoidal function, since inspiration is usually shorter than expiration, and this difference is intrinsic to the respiratory system.

Several decomposition approaches have been considered in the literature, FD is just one of them.
Among the recent AM-FM decomposition proposals are \cite{lin18} and \cite{San18}. \cite{Kow18} gives a useful review of methods and revises several requirements a time series analysis method for an oscillatory signal should satisfy. 

In this paper, a novel decomposition of the signal on AM-FM parametrized components, where the AM is always constant, and FM is modelled using a M\"obius transformation, is presented.
In particular, the individual components are $FMM$ signals, as described in \cite{Rue19}, while the multicomponent signal of order $m$ is denoted as $FMM_m$. 
A  fascinating application, a germ of this work, is \cite{Rue20b} where a physically meaningful wave decomposition for ECG signals is given.

There are significant advantages of the $FMM$ decomposition approach as against its competitors for modeling periodic or quasi-periodic signals. First, the simple parametric formulation that, in particular, enables rigorous and parametrized definitions for basic elements. Second, the interpretability of the parameters and their flexibility to describe and differentiate a variety of wave patterns. And third, the accuracy of the estimators and the robustness against noise.
\\

\begin{figure}[!ht]
	\centering
	\includegraphics[width=0.9\textwidth]{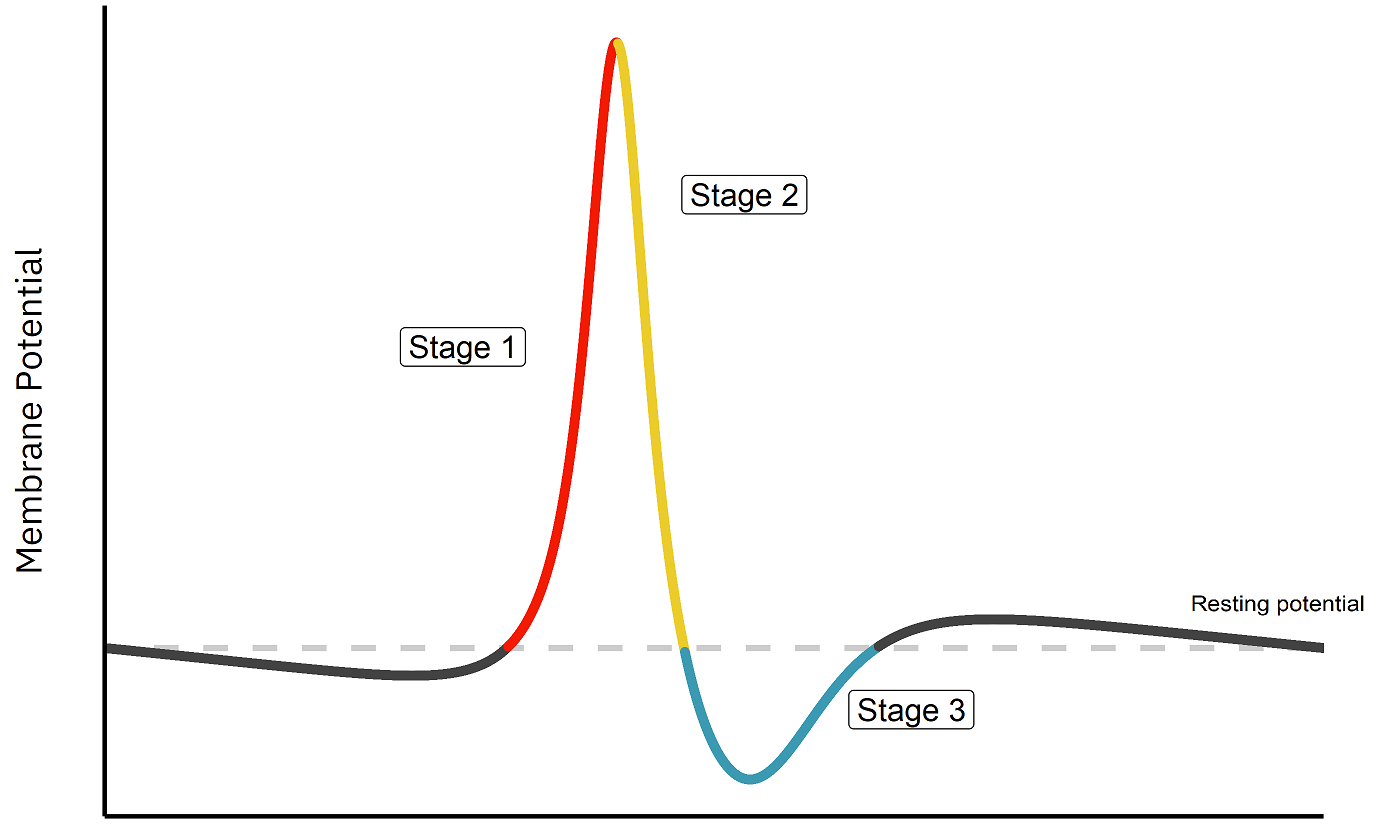}
	\caption{A typical  Action Potential curve and its phases. The signal has been generated  using an  $FMM_2$ model.}
	\label{f:Figure00}
\end{figure}

While this paper addresses a broad audience of data analysts, it is particularly aimed at the mathematicians and statisticians  who will most value the strengths of the model from a theoretical point of view, in addition to the applied one.
Many properties of the model are rigorously described.
In particular,  the  Analytic Signal (AS) associated to the $FMM_m$ model and other essential elements are derived. In addition, a \textit{Dominant Phase} definition is presented which has interesting properties. Moreover, the $FMM_m$ signal is characterized as the solution to a system of differential equations; while, on the statistical side, an estimation algorithm is given and such properties as consistency and accuracy are shown.
Regarding applications, here we deal with problems in Neuroscience, which are also interesting for researchers in electrophysiology and biology. Specifically, we deal with  Action Potential curves (AP) that measure the fluctuation of the potential of a neuron; that is, the difference between the electrical potential inside and outside the cell due to an external stimulus. The AP, which describes the system for about a milisecond, starts from a resting potential of aproximately -70mV  and has several stages. At Stage 1 (Depolarization)  the voltage rises, at Stage 2 (Repolarization) the voltage falls, and at Stage 3 (Hyperpolarization) the negative voltage returns to the resting potential level. If the depolarization is large enough, the cell spontaneously spikes and then goes to a refractory period, during which the cell cannot spike. The typical shape of an AP with a single spike is shown in Figure  \ref{f:Figure00}. For a short introduction see \cite{Rag19}. 
For researchers in the subject, it is of critical importance to extract features of the spike generation of individual neurons, among them, the location and shape of the spike.  These characteristics are important to determine the cell types and their functions and to help us to understand the physiological process, see \cite{Men12} or \cite{Tra19}, among others. 
It can  be said that  Mathematical modeling and analysis of waveform sequences have been one of the central problems in the field of computational Neuroscience.
\\

The description of a dynamic process by Ordinary Differential Equations (ODE) is a traditional approach of analysis. Many models have been described in the literature (see \cite{Tee18} and references therein), the Hodgkin and Huxley model, firstly presented in \cite{Hod52}, is the most widely studied as it has served successfully to study the  bio-electrical activity from different organisms. The family of leaky-integrate-and-fire (LIF) models have also recently become very  popular in the literature (\cite{Tee18}, \cite{Ger14}). 
While these models have biological, physical and chemical foundation, they require a precise measurement of the studied neuron to adjust the parameters and are useful only in controlled experiments. Besides, neuronal activity is often noisy and non stationary across time, which makes the problem of extracting features significantly challenging. Flexible and simple approaches such as the $FMM$ that account for the noise and parametrically describe the oscillatory signals, are suitable to tackle the problem.
\\

In this paper, the potential of  the $FMM$ approach to model AP signals is illustrated using real signals from the Allen Cell Types Database (ACTD) (\url{http://celltypes.brain.map.org}). This database is freely available and has been the reference data for many authors (\cite{Tee18} and references there in). The $FMM_m$ model is compared with two widely used approaches as are the FD and the Spline model. The $FMM_m$ outperforms both of them. Furthermore, across the paper, several properties of the $FMM$ model will be included to construction and analysis of \textit{ Phase Response Curves} (PRC). PRC  describes the variation of quantities (often the phase) of the system in response to perturbations or stimulus. 
\\

The rest of the paper is as follows; Section 2 revises some basic elements of oscillatory systems and Section 3 presents the $FMM$ model and the mathematical and statistical properties. In Section 4, the contributions of the methodology to  the study of AP and PRC curves in Neurosciences are explained and the results from numerical studies with real AP data are shown. Finally, a brief discussion is included in Section 5 and the proofs of theoretical results in the appendix.

\section{ Basic elements in oscillatory systems.}

Different types of variables or signals are defined in periodic oscillatory systems, those applying directly to the motion and usually observable, as the membrane voltage, and those describing the periodic nature of the motion: amplitude, period, and phase, which are not observable directly. The period is the time taken for an oscillating system to return to its initial position, which we assume is known and fixed. The period of the oscillation is normalized here to be $2\pi$.
On the other hand, the phase it is the most elusive of these quantities but a fundamental one, as is the key to describing variations among signals. Hence having a proper definition of phase is essential. 

Moreover, it is generally accepted that for a given oscillatory phenomenon, there exists an underlying complex-valued signal: $S(t)=\mu(t)+i\nu(t)$, $t\in [0,2\pi]$. However, there is no unique way to derive the complex signal when only the real signal is known. The AS approach, the most extended in the literature,
is briefly presented below, along with other elements as the \textit{phase space} and the \textit{periodic orbit}.

\subsection{ FD, HT and AS.}

For many authors, the FD is one of the most important mathematical tools in signal analysis. The FD is the representation of a real signal as a sum of components, as follows:

\begin{equation*}
\mu(t)=a_0+\sum_{k=1}^{\infty}a_k\cos(k\omega_0t)+b_k\sin(k\omega_0t)
\end{equation*}
Besides,  the Hilbert Transform (HT) is considered one of the most critical operators in mathematical analysis that we define below to  facilitate the reading.
\begin{Def}{HT on the real line:}
	\label{def:HT}
	Let  $f \in L^p(\mathbb{R}),$ $1 \leq p < \infty,$ the HT of $f$ on the real line is defined by,
	
	\begin{equation*}
	HT(f(t)) = p.v. \frac{1}{\pi}\int_{-\infty}^{\infty}\frac{1}{t-x}f(x)dx,
	\end{equation*}
	where $p.v.$ denotes the principal value singular integral.
\end{Def}
Finally, the AS associated to a real signal is defined as follows,
\begin{Def}
	
	\label{def:AS} Analytic Signal representation of $\mu(t)$ .

	$S(t)=\mu(t)+ i\nu(t)$, where $\nu(t)=HT(\mu(t))$.
	
\end{Def}

The AS was first defined by \cite{Gab46} as that complex signal, underlying an observed real signal, constructed with the HT. AS has interesting properties and researchers often assume that the underlying complex signal associated to an oscillatory process is an AS, which simplifies the analysis (see \cite{Pic97}, \cite{San15}). 


Given a complex signal, $S(t)=\mu(t)+ i\nu(t)$, it can be expressed as: $A(t)e^{i\phi(t)}$ where,
\begin{equation}\label{eq:phi}
\phi(t)=2\arctan\left(\frac{\nu(t)}{\mu(t)}\right) ;   A(t)=\sqrt{\mu(t)^2 + \nu(t)^2}.
\end{equation}
$A(t)$ and $\phi(t)$ are called the signal's \textit{Instantaneous Amplitude} (IA) and \textit{Instantaneous Phase} (IP), respectively. The derivative of $\phi(t)$ is known as \textit{Instantaneous Frequency} (IF),which is expected to be positive in applications, as argued for instance in \cite{San15}. Hence, the AS is not always interpretable
in a way which is meaningful and representative of physical
phenomena. In particular for multicomponent signals (\cite{Boa16}). Nevertheless, the signal could be modeled as a weighted sum of component signals,  as in \cite{San18},  in which case the problem is that the decomposition is not unique. In order to get interpretable results, the role of each of the components should be identified. This question is dealt with later in the paper.

\subsection{The phase, the phase space and the periodic orbit.} 

There are multiple definitions of phase in the literature that may lead to contradictory results.  A simple property, remarked by \cite{Win01}, is that every point on the oscillation can be uniquely described by a phase. Hence, for many authors, the phase has a natural definition as the time along the cycle  (\cite{Win01}, among others). Besides, for signals such as  $\mu(t)=A\cos(\phi(t))$ where $\phi(t)$ is a periodic function, $\phi(t)$ is also an interesting phase definition. A popular approach, adopted by some authors such as \cite{Den16}, \cite{Opr17}, and \cite{Car19}, is to use the IP associated with the AS approach, defined in (\ref{eq:phi}).

The ambiguity on phase definition is well explained in \cite{Osi03}, \cite{Cha06}, and \cite{Fre18} where other alternatives are also provided.

The phase definition, to be useful in practice, should be also applied to the underlying complex signal, and reciprocally, the representation on a complex plane is essential to derive a proper phase definition. Hence, the concept of \textit{phase space}, a space in which all possible states of a system are represented, with each possible state corresponding to one unique point, is also crucial in dynamic systems.

The degrees of freedom or dimensionality of a dynamic system is the number of variables governing the state of the system at time \textit{t}. The \textit{phase space} has the same dimension as the degrees of freedom of the system and often is two, which cases it is called a phase plane.

In classical mechanics and other fields, the phase space is obtained by plotting the positions against the velocity (\cite{Car18}). The phase plane associated with an AS is obtained by plotting the real signal $\mu(t)$ against $HT(\mu(t))$, and the angle at a given point is the IP.
\\

The system's evolving state over time traces a trajectory through the phase space. The trajectory of a periodic system, the image of the periodicity interval in the state space, is a closed curve called the \textit{periodic orbit} or cycle. Given a closed curve and a point in the phase space, the \textit{winding number} is the integer representing the total number of times that curve travels counter-clockwise around a point in the interior. The maximum value of the winding number can be interpreted as the number of oscillations within a period. A signal, typically monocomponent, with only one oscillation, describes a closed orbit with maximum winding number of 1 (\cite{Kra12}).

If a center point with maximum winding number is found, then the angle phase definition with respect to that point is an admissible phase definition. The main drawback is that very often such a point is not easy to find. Some examples are given in the next section.     

\section{ The $FMM_m$ model: a novel decomposition approach}

Oscillatory signals are defined in the time domain and, without loss of generality, it is assumed that the time points are in  $[0,2\pi]$. In any other case, transform the time points $t' \in [t_0,T+t_0]$ by $t=\frac{(t'-t_0)2\pi}{T}$. 
In the following, oscillations are also referred to as waves. 
\subsection{Definition and statistical properties.}
Let, $\boldsymbol{\upsilon}=(A,\alpha,\beta,\omega)'$ be the four-dimensional parameters describing a single FMM signal, defined as the following \textit{wave}: $	W(t,\boldsymbol{\upsilon})=A\cos(\phi(t,\alpha,\beta,\omega)),$ where $A$ is the wave amplitude and,
\begin{equation}\label{phi}
\phi(t,\alpha,\beta,\omega)=
\beta+2\arctan(\omega\tan(\frac{t-\alpha}{2}))
\end{equation}
is the wave phase.
The additive $FMM_m$ model is defined as a parametric additive m-component signal plus error model as follows;

\begin{Def}{ \textit{$FMM_m$ model}} \label{eq:mFMM}
	
	For the observations $t_1<...< t_n$,
	\begin{equation}\label{eq:mod}
	X(t_i)= \mu(t_i,\boldsymbol{\theta})+ e(t_i); 
	\end{equation}
	\quad	where, $
	\mu(t_i,\boldsymbol{\theta})= M+ \sum_{J=1}^{m} W(t_i,\boldsymbol{\upsilon_J})$, and,
	
	\begin{itemize}

		\item  $\boldsymbol{\theta}=(M,\boldsymbol{\upsilon_1},...,\boldsymbol{\upsilon_m})$ verifiying:
		\begin{itemize}
			\item $ M \in \Re $; $\boldsymbol{\upsilon_J} \in \Theta_J = \Re^+ \times [0,2\pi] \times [0,2\pi] \times[0,1]$; $J=1,...,m$,
			\item $\alpha_1 \le \alpha_2 \le....\le \alpha_m \le \alpha_1$
			
			\item $A_1 =\max_{1 \le j \le m}A_j $ 
			
		\end{itemize}
		\item $(e(t_1),...,e(t_n))' \sim N_n(0,\sigma^2\boldsymbol{I})$, 
	\end{itemize}
	\vspace{1mm}
\end{Def}
The identifiability of the model parameters is guaranteed by including the artificial restrictions above. The papers by \cite{Rue19} and \cite{Rue20b} considering particular cases of this model, show the broad type of signals that the model represents and provide parameter interpretation as well as some basic properties. 
\\

\textbf{Maximum Likelihood Estimator (MLE)}

The MLE of the $FMM_m$ model parameter are the solutions to the optimization problem:
\begin{equation}\label{eq:MLE}
\hat{\boldsymbol{\theta }}=\argminA \limits_{\boldsymbol{\theta} \in
	\Theta}{\sum\limits_{i=1}^n(X(t_i)-\mu(t_i,\boldsymbol{\theta}))^2},
\end{equation}
\vspace{1mm}
where $\Theta$ refers to the parameter space for $\boldsymbol{\theta} $, a subset of  $R^{4m+1}$ given by $\Re \times \Theta_1  \times ...\times \Theta_m$  plus the restrictions.
When the true parameter configuration verifies $\alpha_J \in (0,2\pi),\beta_J \in (0,2\pi),w_J>0 ; J=1,...,m$, the standard regularity conditions on the response function are verified for $FMM_m$ and well known results in nonlinear normal regression guarantee the consistency and asymptotic normality of the MLE estimators.
The main pitfall is how to find the MLE.\\


A backfitting algorithm, Algorithm \ref{alg:1} below, is proposed to solve the optimizing problem (\ref{eq:MLE}),  which at each step,  fits an $FMM_1$ model to the residue using the algorithm designed by \cite{Rue19}. This is repeated until the difference between the variability explained by the model in two consecutive steps is less than a constant $C$. $C$ depends on the experiment and on the researcher.

A measure of the variance proportion explained by the model is defined as follows: 
\begin{equation}\label{R2}
R^2=1-\frac{\sum_{i=1}^n (X(t_i)-\mu(t_i,\hat{\boldsymbol{\theta}}))^2}{\sum_{i=1}^n (X(t_i)-\overline{X})^2}
\end{equation}
being $n$ the number of observed values.

\begin{algorithm}
	
	\small{
		
		\caption{MLE $FMM_{m} (\boldsymbol{\theta})$ estimation}
		\label{alg:1}
		\begin{enumerate}
			\item \textbf{ Initialize} for $J=1,\dots,m$:
			$$M= \frac{1}{n}\sum_{i=1}^n X(t_i); A_J=0,\alpha_J=5,\beta_J=\pi,\omega_J=1; J=1,\dots,m$$
			
			\item \textbf{Do until $R^2$ increases less than $C$ },
			\textbf{For each  $J; J=1,...m$:}
			\begin{enumerate}[2.1]
				
				\item 	$ \hat{\boldsymbol{\upsilon}}_J, \hat{M}  \leftarrow  \argminA \limits_{\boldsymbol{\upsilon}_J \in \Theta_J; M \in \Re}{\sum\limits_{i=1}^n(X(t_i)- \sum_{I \neq J} \hat{W}(t_i,\boldsymbol{\upsilon}_{I}) - M- W(t_i,\boldsymbol{\upsilon}_{J}) )^2} $ 
				
				\item \text{Order the components using $A_1=max_{1 \le j \le m}A_j$ and  $\alpha_1 \le...\le \alpha_m \le \alpha_1$}
				
				\item $\mu(t_i,\hat{\boldsymbol{\theta}})=\hat{M}+\sum_{J=1}^{m}W(t_i,\hat{\boldsymbol{\upsilon}}_{J})$ 
				
				\item Calculate $R^2$
			\end{enumerate}
			\vspace{-1.5mm}
		\end{enumerate}
	}
\end{algorithm}


Success, in terms of convergence to the MLE from a given starting value, is not initially
guaranteed, although the solution converges in probability to a local minimum.
Our experience fitting $FMM_1$ to real and simulated data indicates that the failure of convergency does not likely happen. Moreover, the excellent performance of the backfitting algorithm has been shown with the simulations results for the $FMM_5$ model describing the ECG signal in \cite{Rue20b}. In this paper, we have checked that this is also true in the particular case of $FMM_3$ models describing real action potential curves.
\\


The likelihood-based analysis of the $FMM_m$ model would benefit from the ability to conduct hypothesis testing problems or derive confidence intervals. Specifically,  assuming the $FMM_1$ model, both hypothesis tests on \textit{arrhythmicity} and the \textit{sinusoidal shape} are defined parametrically, see \cite{Rue19}. Moreover, other interesting hypothesis testing problems can be defined depending on the problem at hand.  
While the parameters that describe the hypothesis are conveniently chosen, it is straightforward to develop the likelihood ratio test and confidence intervals using such  standard methods as bootstrap.
\\

For the rest of the paper, we refer to the $FMM$ model or $FMM$ signal depending on  whether (\ref{eq:mod}) or  $\mu(t,\boldsymbol{\theta})$ is considered, that is the noise is considered or not. In addition, the dependence of signals, waves, phases, and models on the parameters is omitted when no confusion. Specifically, $W_J(t)=W(t,\boldsymbol{\upsilon}_J)$, $\phi_J(t)=\phi(t,\alpha_J,\beta_J,\omega_J)$, and  $\mu(t)=\mu(t,\boldsymbol{\theta})$.
\vspace{-2mm}
\subsection{New theoretical properties}
Without loss of generality, we assume for an $FMM_m$ signal that $M=0$ for the discussion in this section, as $M$ can always be assigned to the component 0 where $A_0=M$ and $\phi_0(t)=0$.

\vspace{-1.5mm}
\subsubsection{The AS associated to an $FMM_m$ signal}
In general, given a real signal $\mu(t)$, the associated AS does not have a closed expression even when the signal has a simple expression as $\mu(t)=B\cos(\psi(t))$, as could be expected. Examples in \cite{Pic97} illustrate this statement. However, for $FMM_m$ signals, the AS can be easily derived analytically, as shown in Theorem \ref{pr:as}.

\begin{Theo} \label{pr:as}
	Let $\mu(t)$ be an $FMM_m$ signal,  the AS associated is $S(t)=\mu(t)+i\nu(t)$, where,
	$\nu(t)=\sum_{J=1}^{m} A_J\sin(\phi_J(t)) $.
\end{Theo}

The proof follows, taking into account that $HT(\sum_{J=1}^{m} W_J(t))=\sum_{J=1}^{m} HT(W_J(t))$ and that $HT(W_J(t))=A_J\sin(\phi_J(t)) $, the latter is shown in \cite{Rue19}.\\

Now, the AS phase is easily derived as the angle of the vector  $(\mu(t),HT(\mu(t))$ with respect to $(0,0)$. Moreover, with some analytical work, the IF can also be derived, and then it is not difficult to find examples where the IF is negative. 

\cite{Wei98} gives conditions under which a multicomponent signal, such the $FMM_m$ signal, has an IF valid interpretation. Specifically, for the $FMM_2$ model,  a necessary condition for IF be interpreted as a non negative weighted average of the IF's of the two components, taking into account that $A_1>A_2$, is:
\begin{equation*}
\frac{A_2}{A_1} \ge -\cos(\phi_1(t)-\phi_2(t)).
\end{equation*}

\begin{figure}[!ht]
	\includegraphics[width=1\textwidth]{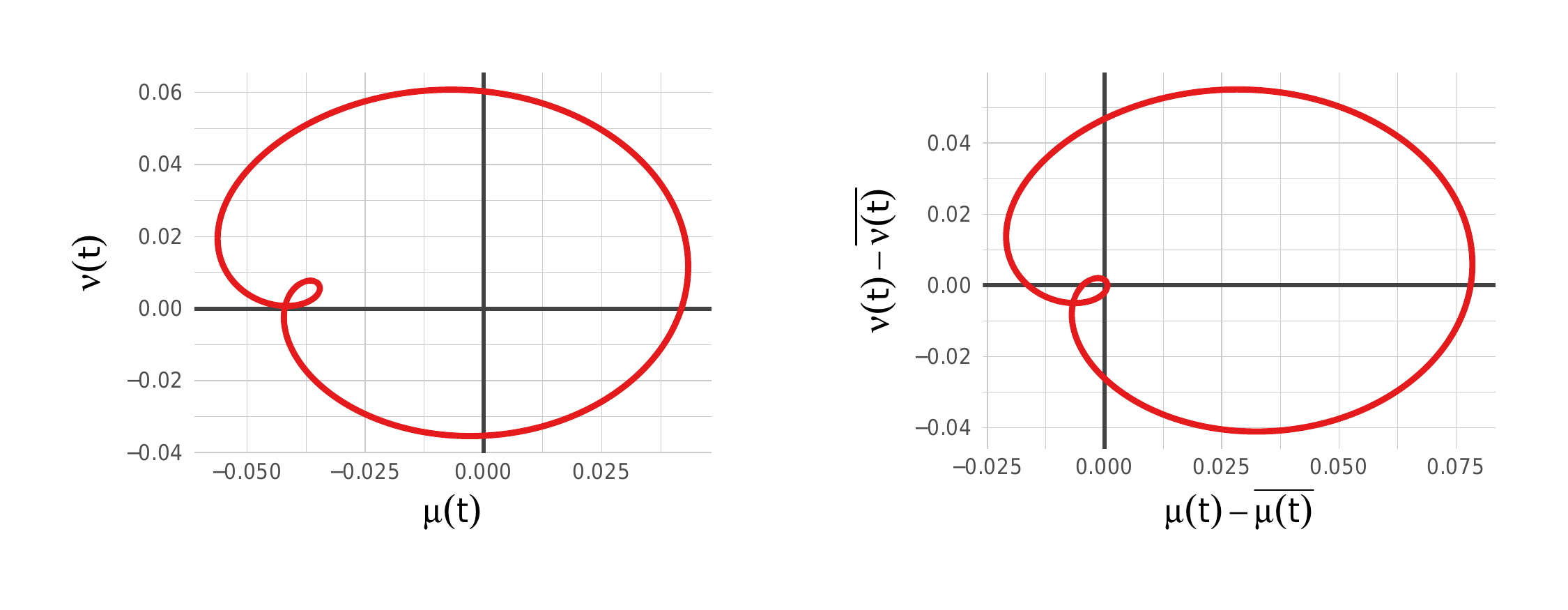}
	\caption{ The phase spaces, with the real and imaginary signals defining the AS, are plotted  for $\mu(t)$ (left) and $\mu(t)-\overline{\mu(t)}$ (right) respectively. where, $\mu(t)$  is the signal in Figure \ref{f:Figure00} and $\overline{\mu(t)}$ is the mean.}
	\label{f:Figure1}
\end{figure}

\begin{figure}[!ht]
	
	\includegraphics[width=1\textwidth]{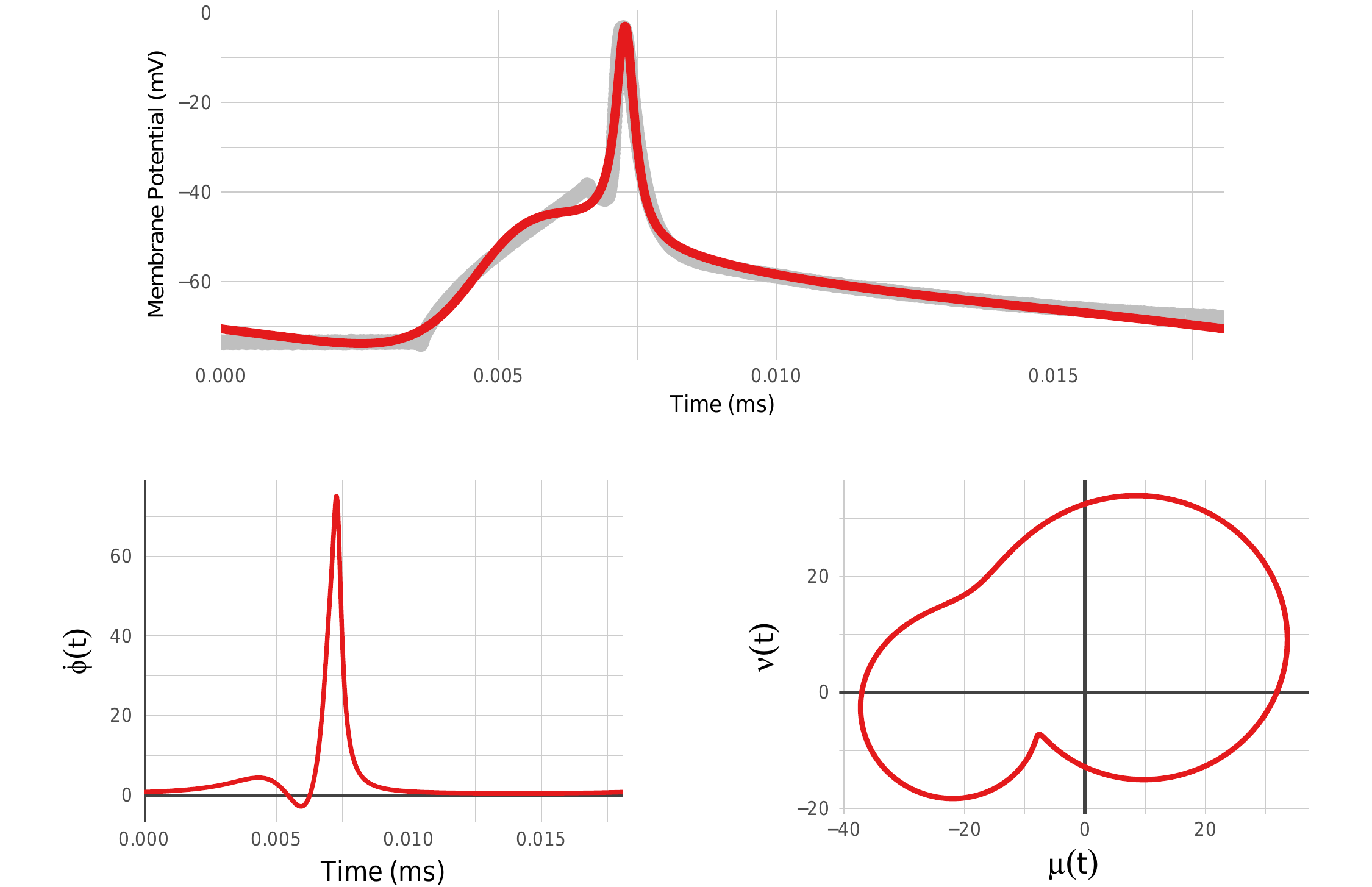}
	\caption{The plot on the top is a real oscillatory signal (grey) from the ACTD (experiment 486754703, sweep 17)  and the estimated $FMM_2$ signal in red ($\mu(t)$). The plots on the bottom are $\mu(t)$ against $HT(\mu(t))$ (right),  and $t$ against $\dot{\phi}(t)=\frac{\partial \phi (t)}{\partial t}$(left). }
	\label{f:Figure2}
\end{figure}

For $FMM_m$ with $m>2$, the conditions for a valid IF are more demanding. \cite{Cha06} and \cite{Fre18} show scenarios, like those in \cite{Wei98}, where the AS fails, corresponding to signals with more than one dominant oscillation. In this section, two examples are shown that illustrate how AS fails even in scenarios where there is an apparent single oscillation. First, in Figure \ref{f:Figure1}, the same signal in Figure \ref{f:Figure00} is considered, the phase space for that signal in Figure \ref{f:Figure00}(left) and the phase space for the same but centered signal in Figure \ref{f:Figure00}(right), are plotted; quite different AS phases are defined from these two signals describing the same system, as the center point (rotation center) is different. The second example is shown in
Figure \ref{f:Figure2}, where the real signal (experiment number  486754703, sweep 17 from ACTD) is analysed; a $FMM_2$ model is fitted to the observed data (top), and the associated phase space (bottom right) and IF (bottom left) are plotted.
\\

Therefore, even when the signal  exhibits a single dominant oscillation, the AS phase definition could fail and an alternative phase definition is needed; even more considering that the calculation of the phase and IF are highly susceptible to background noise. 

The simple analytical expression of the $FMM$ model facilitates a proper and robust phase definition that is presented below as the \textit{Dominant Phase}.


\subsubsection{The Dominant phase}
In some applications, a noticeable characteristic of the signal is the existence of a dominant component, as is the case of AP from single neurons where the dominant component corresponds to the moment when the neuron spikes (\cite{Wei98}). The dominant component amplitude is expected to be much larger than those of other components, in such cases. Therefore, the signal phase, IA, and IF are approximately identical to those of the dominant component. 

These statements are the basis of the definition of the \textit{Dominant Phase} (DP), for $FMM_m$ signals, Definition \ref{def:phi} (a) and the \textit{Dominant Peak Time} (DPT),  Definition \ref{def:phi} (b). Moreover, the definitions for the \textit{Dominant Instantaneous Frecuency} (DIF) and the \textit{Dominant Instantaneous Amplitude }(DIA), are $\dot{\Phi}_D(t)=\frac{\partial\Phi(t)}{\partial t}$ and $A_1$, respectively. 
In the simple case,  $m=1$, the DP coincides with IP from the AS.

\begin{Def}{The dominant phase and the dominant phase peak} \label{def:phi}. Let  $\mu(t)$ be an $FMM_m$ signal where $A_1=max_{ 1 \le J\le m}A_J$
	\begin{enumerate} [(a)]
		\setlength\itemsep{-0.4em}
		\item   The dominant phase is defined as: $\Phi(t)=\phi_1(t)-\beta_1$
		\item  The  dominant peak time is defined as: $\Phi_{peak}=2 \arctan (\frac{1}{\omega_1}\tan(\frac{-\beta_1}{2}))$	
	\end{enumerate}	
\end{Def}

Some interesting properties are shown in Proposition \ref{def:pro}.  First, it is shown that $\Phi(t)$ increases monotonically with time, which makes the formulation physically interpretable. Moreover, the derivatives of $\Phi(t)$ and  $\Phi_{Peak}$ with respect to the parameters are given because can be useful to construct PRCs. We will return to this question in section \ref{Neuro}.
\begin{prop} 
	\label{def:pro} 
	Let  $\mu(t)$ be an $FMM_m$ signal and $\Phi(t)$ as above, then :
	\begin{enumerate}
		\item 
		\begin{enumerate}[(a)]
			\item $\frac{\partial\Phi(t)}{\partial t}=\omega_1+\frac{1-\omega_1^2}{2\omega_1}(1-\cos(\Phi(t))$
			
			\item $\frac{\partial\Phi(t)}{\partial \alpha_1}=-[\omega_1+\frac{1-\omega_1^2}{2\omega_1}(1-\cos(\Phi(t))]$
			
			\item $\frac{\partial\Phi(t)}{\partial \omega_1}
			= \frac{1}{\omega_1}\sin (\Phi(t))$
		\end{enumerate}
		\item
		\begin{enumerate}[(a)]
			\item $\frac{\partial \Phi_{Peak}}{\partial \beta_1}=-[\frac{1}{\omega_1}+\frac{1-\frac{1}{\omega_1^2}}{2\frac{1}{\omega_1}}(1-\cos(\Phi_{Peak})]$
			
			\item $\frac{\partial\Phi_{Peak}}{\partial \omega_1}=-[\frac{1}{\omega_1}\sin (\Phi_{Peak})]$
		\end{enumerate}
	\end{enumerate}

\end{prop}
The proof is deferred to the appendix.
\\

Note that the derivatives in Proposition \ref{def:pro} are formulated as functions of $\Phi(t)$ and  $\Phi_{Peak}$, respectively, an not only as function of time, which is useful in applications. More specifically, it is relevant for real practice to note that also  $-\frac{\partial\Phi(t)}{\partial \alpha_1}$, $-\frac{\partial \Phi_{Peak}}{\partial \beta_1}$ and $-\frac{\partial \Phi_{Peak}}{\partial \omega_1}$ are non negative functions.
\vspace{-2mm}
\subsubsection{The ODEs representing the $FMM_m$ signals}

Dynamical models describing the state of a system are frequently formulated in terms of ODEs, very often in Neuroscience. The derivation of  the ODE representation of the $FMM$ signal is interesting to compare with alternative models and to show other aspect of the dynamics that the signal describes.

The problem is known as \textit{inverse problem} for ODEs. Given a function signal, find an ODE  $ f(x,\dot{x},\ddot{x},..,t)=0$,  $\dot{x}=\frac{\partial x(t)}{\partial t}$, that admits that signal as a solution. The results in this section are inspired by the work of \cite{Wig15}, where the conditions under which a periodic signal  can be represented by an ODE of order \textit{k} are derived. Specifically, it is shown that the minimal order depends on the minimal dimension in which the stable orbit of the system does not intersect itself. For an $FMM_1$ signal, this dimension is two, as Theorem \ref{th:ode} shows. 

Moreover, using a change of variable, we derive a second order ODE associated to the DP that describe phase dynamics (phase model).

\begin{Theo}
	
	Let  $\mu(t)$ be an $FMM_1$ signal with  $\omega_1 >0$ and  $z(t)=\tan(\frac{\Phi(t)}{2}),$ then
	
	\label{th:ode}
	\vspace{-1mm}
	\begin{enumerate}[(a)]
		
		\item  $\mu(t)$ is the solution to the following equation: \\
		\vspace{-1mm}
		$ \ddot{x}(t)=-x(t)\phi_1(t)+\dot{x}(t)\frac{\ddot{\phi_1}(t)}{\dot{\phi}_1(t)}$

		\item
		$z(t)$ is the solution to the following equation: \\
		\vspace{-1mm}
		$ \dot{x}(t)=\frac{\omega_1}{2}+\frac{1}{2\omega_1}x^2(t)$
	\end{enumerate}
	
\end{Theo}

The proof is deferred to the appendix.
\\

Furthermore, a system of ODE's with an $FMM_m$ as a solution can be derived, as is done for instance in \cite{Wig05}, using Theorem \ref{th:ode}) and the additive structure of $FMM_m$.
However, the minimum order of the ODE representing an $FMM_m$ model can not be predicted in advance, as it depends on the parameter configuration.

\section{Applications in Neuroscience} \label{Neuro}
Neuroscience can be defined as a multidisciplinary branch of biology that combines physiology, anatomy, molecular biology,  mathematical modeling, and psychology to understand the nervous system. We deal here with neuron cells.

Much of the mathematical treatment of the nervous system has its roots in the theory of ODEs. It  has been promoted for many years in the work of \cite{Win01}, \cite{Hol00}, \cite{Kop86}, \cite{Erm81},  and  \cite{Izh07},  to name just a few. For a survey, we refer the reader to the book by \cite{Erm10}. 
Models that describe nervous signals can be classified as empirical or mechanistic; the former attempts to describe spiking output and are based on direct observation, while the latter attempts to describe physiological features and are based on an understanding of the behavior of a system's components.  The $FMM_m$ model is of the former class and the Hodking and Huxley  of the latter. The empirical models are particularly useful to analyse \textit{in-vivo} data.

The estimation of the phase and other quantities associated with the system depend on the specific approach. In particular, a critical curve to be estimated from experimental data is the PRC, also known as phase resting curve or phase sensitivity curve. 
There is no consensus on phase definition, and nor is there in the estimation of the PRC from experimental data. The subject has received much attention in the literature as PRCs are used for multiple proposes;  for instance see \cite{Sch11}. More specifically, \cite{Opr17} , \cite{Shi19} and \cite{Ros18}, propose using the AS to estimate PRCs.

The $FMM$ solves the construction of PRCs satisfactorily. Let us assume that a perturbation of a system can be represented by a change in one of the parameters; hence, the PRC can be obtained by estimating the derivative of the DP with respect to each of the parameters. Alternatively, it could be also interesting to measure changes in DPT, the PRCs could then be derived by calculating the relative changes in the DPT.  
The two types of PRCs documented in the literature  are:  Type-I (positive, phase advanced) and Type -II (positive and negative, advanced, and delay phase), see \cite{Koo09}.
Proposition \ref{def:pro} shows how the two types of the functional forms, Type I and Type II, arise depending on which parameter is changing.

Furthermore, if the derivative with respect to \textit{t} is considered to calculate the PRC,  the ODE derived in  Theorem \ref{th:ode} (b) shows that the model associated to the DP is closely related to  the \textit{theta model}, also known as the the Ermentrout-Kopell canonical model (\cite{Erm96}). Specifically, the  $FMM$ model is equivalent to this latter model, when $\omega_1=0.5$, assuming $\omega_1^2=I$, where $I$ is the stimulus intensity.
Therefore, when the DP is considered,  only phase advanced are produced by a perturbation. Accordingly, the classification in Type I or Type II models depends on when the DP is adopted or not; which, in turns, depends on the user, the number of components and the variability explained by the dominant component.

Whatever the definition of PRC is chosen, the $FMM$ approach simplifies the estimation process because the PRC is formulated parametrically.

Regarding the AP curve, Hodking and Huxley and other ODE models have been extensively used to fit AP from in vitro data. However, the models are not useful for experimental or in vivo data, as is explained and illustrated for instance in \cite{Nau06}. The FMM approach achieves a quasi perfect fit for different AP patterns, as can be seen with the numerical analysis below.
\\

Other interesting applications are mentioned in the discussion section.

\subsection{Analysis of AP from ACTD}

The ACTD includes morphological and electrophysiological data collected from individual human or mouse recordings of high temporal resolution time series of membrane potential.
The APs from the first 500 recorded neurons, using the short square stimulus and the lowest stimulus amplitude generating a spike, have been analysed.

The time needed by the neuron to spike following the application of the stimulus ($d$) is used to delimit the segment containing the AP, which is defined as $[t_S-2d, t_S+3d]$ with $t_S$ denoting the time of the spike. This uneven cut is done to capture the asymmetry of the AP, as the depolarization happens much faster than the rest of its stages. The number of observations, depends on the experiment, ranging from 500 to 4500. Neurons from two species, human and mouse are analysed. According to the dentrite type, neurons are classified as inhibitory or excitatory. 18\% are human neurons (22\% of them inhibitory) and 82\% mouse neurons (49\% of them inhibitory).

\begin{figure}[hbtp]
	\centering
	\includegraphics[width=1\textwidth]{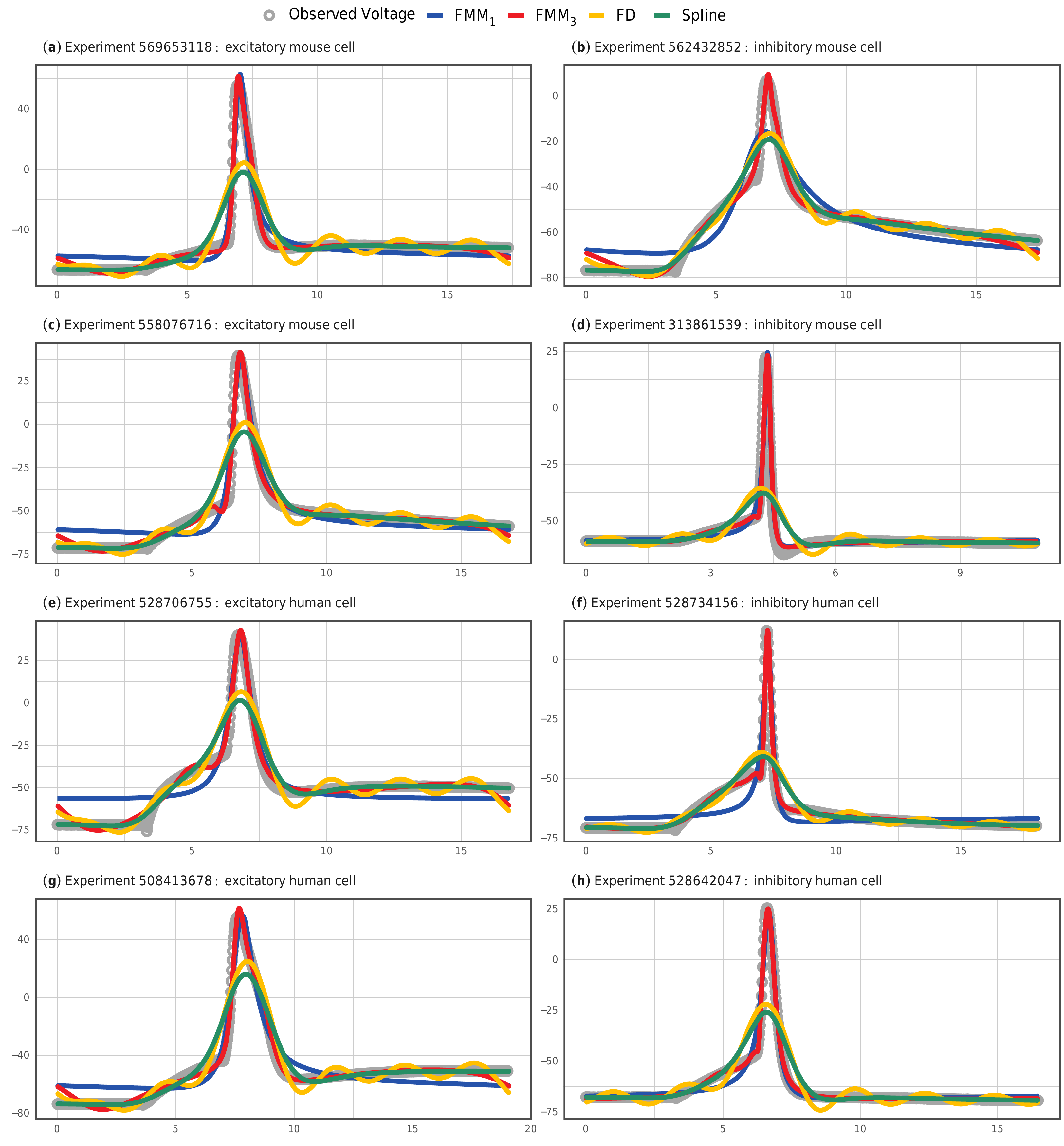}
	\caption{AP extracted from different experiment from ACTD along with the fitted signal using Spline with 13 df (green), FD with six harmonics (yellow), $FMM_1$(blue), and $FMM_3$(red).}
	\label{f:Figure04}
\end{figure}

The $FMM_1$, Spline, FD, and $FMM_3$ models have been fitted to the signals. The Spline and FD models fitted are comparable to the $FMM_3$ in complexity. Therefore, as the $FMM_3$ model has 13 parameters, a 13 df (degrees of freedom) Spline and an FD with six harmonics, have been considered. Figure \ref{f:Figure04} shows the AP for inhibitory and excitatory neurons from humans and mice with different patterns.

Consider the measure $R^2$ defined in (\ref{R2}) as a measure of the goodness of fit of a model.

Among the four models fitted to the data, the highest $R^2$ is always that of $FMM_3$. In most cases, $R^2$ is also higher for  $FMM_1$  than for the Spline or FD models. Table \ref{table:R2_Table} gives $R^2$ means and standard deviations across types and species.

As Figure \ref{f:Figure04} and numbers in Table \ref{table:R2_Table} illustrate, most signals are quite well represented with an $FMM_1$ model, in particular inhibitory neurons.   The latter is an interesting fact as $FMM_1$ is a much more simple model with 5 df, as against the 13 df of the other three models.

\begin{table}
	\caption{\label{table:R2_Table} $R^2$'s mean (standard deviation) across dentrite type and species.}
	\centering
	\begin{tabular}{l l c c c c}
		\toprule
		\textbf{Dendrite Type} & \textbf{Species} & $\mathbf{R^2_{FMM_3}}$ & $\mathbf{R^2_{FMM_1}}$ & $\mathbf{R^2_{Spline}}$ & $\mathbf{R^2_{FD}}$ \\
		\midrule
		\textbf{Inhibitory} & \textbf{Human} & 0.992&0.902&0.675&0.652\\
		& & (0.004)&(0.064)&(0.083)&(0.092)  \\
		& \textbf{Mouse} &0.992&0.874&0.736&0.703\\
		& & (0.005)&(0.058)&(0.084)&(0.091)  \\
		\textbf{Excitatory} & \textbf{Human} &0.981&0.892&0.817&0.802\\
		& & (0.006)&(0.053)&(0.053)&(0.059)  \\
		& \textbf{Mouse} &0.982&0.879&0.819&0.788\\
		& & (0.006)&(0.056)&(0.049)&(0.052) \\
		\bottomrule
	\end{tabular}
\end{table}

\begin{figure}[hbtp] 
	\includegraphics[width=\textwidth]{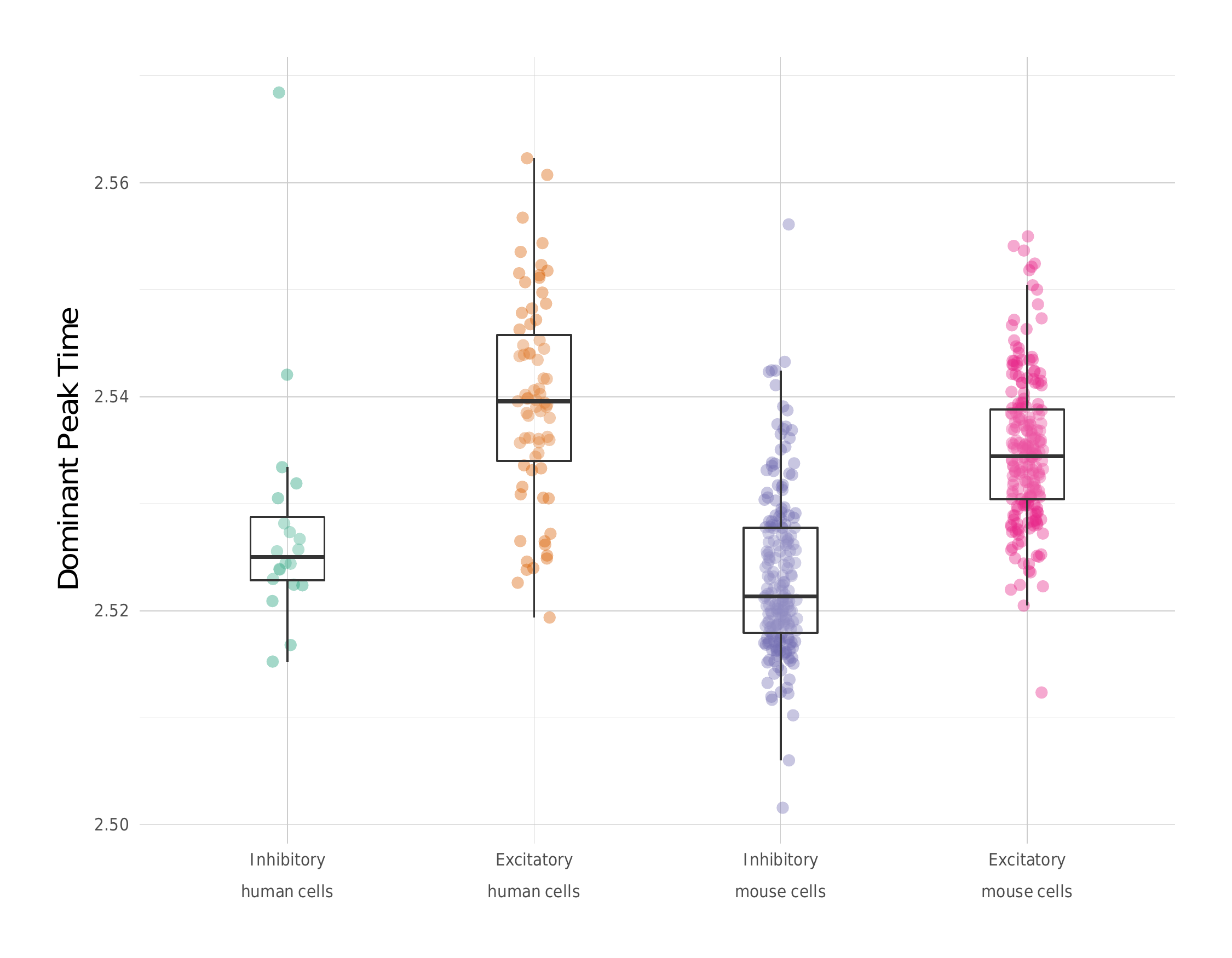}
	\caption{Distribution of the Dominant Peak Time across dendrite type and species.}.
	\label{f:Figure01}
\end{figure}

The ability of the parameters to characterize different transgenic lines or their potential in supervised classification is beyond the scope of this paper and will be part of our future research. An insight of the potential of the $FMM$ parameters to discriminate cell types is given in  Figure \ref{f:Figure01}, which shows how the DPT distribution differs across Species and dendrite types.

\section{Discussion}

In this paper, the $FMM$ approach is presented as a multi-purpose approach with solid mathematical and statistical support. It provides a   decomposition of a periodic signal in several components, with a parametric formulation that facilitates the interpretability and the derivation of essential elements.
Moreover, the ODE representation for the $FMM$ signal captures the dynamics,  and,  on the statistical side, the estimation algorithm and other inference tools allow the analysis of observed signals in the presence of noise.
\\

From the applied side, AP curves from neuron synapses have been analysed using the $FMM $ approach and questions related to the PRC estimation have been addressed. However, many questions still remain open in Neuronal Dynamics. Along with the performance of the proposed PRC estimators in real practice, we can cite, the potential of the model parameters to define synchronization measures (\cite{Ayd13}). 
Moreover, we have focused here on the AP of neurons; however, nerves and muscle and other AP from extracellular recordings are also of great interest. Specifically, three basic waveforms can be defined: monophasic, biphasic or triphasic, based on where the recording electrode is placed (\cite{Rag19}). The $FMM$ parameters can accurately discriminate between these patterns.

Actually, the algorithmic extraction and categorization of the distinct AP is one of the most exciting problems in data analysis in neurophysiology. It is known as \textit{spike sorting}, and has been generating much attention recently in the literature (\cite{Rey15}, \cite{Car18}, \cite{Tee18}, \cite{Sou19}, \cite{Rac20} to mention only a few). The $FMM$ parameters could contribute efficiently to solving the problem of feature extraction in the classification process.
\\

Furthermore, there are multiple questions related to other  electrophysiological signals, such as the EEG signal and other brain signals, where the $FMM$ approach could be useful. More specifically, phase related quantities have been widely used in the analysis of cerebral disorders, as is illustrated in \cite{Sam17} and \cite{Ata09}, among others.
\\

Finally, there are many other fields with a tradition in signal analysis where the $FMM$ as a model or decomposition approach could be useful too, starting by providing a kind of bandwidth filtering.  

\bigskip
\section*{Acknowledgments}
The authors gratefully acknowledge the financial support received by
the Spanish Ministerio de Ciencia e Innovaci\'on and European
Regional Development Fund; Ministerio de Econom\'ia y Competitividad
grant [MTM2015-71217-R and PID2019-106363RB-I00 to CR and YL].

\section{Appendix}

\textbf{Proof of Proposition \ref{def:pro}}

From (\ref{phi}), we have that,
\begin{equation}\label{prueba1}
\tan(\frac{\Phi(t)}{2})=\omega_1\tan(\frac{t-\alpha_1}{2});
\end{equation}

and, using the derivative of the $\arctan$ and the trigonometric equality,  $\cos^2(\theta)=\frac{1}{1+\tan^2(\theta)}$, we have that, \\
$$\frac{\partial\Phi(t)}{\partial t}=\frac{\omega_1\frac{1}{\cos^2{(\frac{t-\alpha_1}{2}})}}
{1+\omega_1^2\tan^2(\frac{t-\alpha_1}{2})}=\frac{1}{\omega_1}\frac{\omega_1^2+\omega_1^2\tan^2(\frac{t-\alpha_1}{2})}
{1+\omega_1^2\tan^2(\frac{t-\alpha_1}{2})}$$.

Now, from (\ref{prueba1}) and the last statement it follows that, 
\begin{equation}\label{prueba1b}\frac{\partial\Phi(t)}{\partial t}=\frac{1}{\omega_1}\frac{\omega_1^2+\tan^2(\frac{\Phi(t)}{2})}
{1+\tan^2(\frac{\Phi(t)}{2})}=\omega_1\cos^2(\frac{\Phi(t)}{2})+\frac{1}{\omega_1}\sin^2(\frac{\Phi(t)}{2}).
\end{equation}
Finally, 1.(a) is the result of applying the trigonometric equalities: $\cos^2(\frac{\theta}{2})=\frac{1+\cos{\theta}}{2}$ and  $\sin^2(\frac{\theta}{2})=\frac{1-\cos{\theta}}{2}$ to the right hand of (\ref{prueba1b}), as follows:
$$\frac{\partial \Phi(t)}{\partial t}=\frac{\omega_1}{2}(1+\cos(\Phi(t)))+\frac{1}{2\omega_1}(1-\cos(\Phi(t)))=\omega_1+\frac{1-\omega_1^2}{2\omega_1}(1-\cos{\Phi(t)})$$

Proposition \ref{def:pro}, 1.(b) is proved in a similar way provided that
$\frac{\partial\Phi(t)}{\partial \alpha_1}=-\frac{\partial\Phi(t)}{\partial t}$.
\\

Proposition \ref{def:pro}, 1.(c) is also proved  using similar arguments as above and the equality, $\sin(2\theta)=2\sin(\theta)\cos(\theta)$, as follows:
$$\frac{\partial\Phi(t)}{\partial \omega_1}=\frac{2\tan{(\frac{t-\alpha_1}{2}})}{1+\omega_1^2\tan^2(\frac{t-\alpha_1}{2})}=\frac{\frac{2}{\omega_1}\tan(\frac{\Phi(t)}{2})}{1+\tan^2(\frac{\Phi(t)}{2})}=\frac{2}{\omega_1}\sin(\frac{\Phi(t)}{2})\cos(\frac{\Phi(t)}{2})=\frac{1}{\omega_1}\sin(\Phi(t)).$$

In addition, Proposition \ref{def:pro}, 2.(a) and 2.(b) follow in  a similar way and the proofs are left to the reader.
\\

\textbf{Proof of Theorem \ref{th:ode}}.

On the one hand, we have that 
$$\dot{\phi}_1(t)=\frac{\omega_1\frac{1}{\cos^2{(\frac{t-\alpha_1}{2}})}}
{1+\omega_1^2\tan^2(\frac{t-\alpha_1}{2})}= \frac{\omega_1}{\cos^2(\frac{t-\alpha_1}{2})+ \omega_1^2\sin^2(\frac{t-\alpha_1}{2})},$$

which implies that $\dot{\phi}_1(t)>0$ provided that $\omega_1>0$.

On the other hand, let  be $x(t)=A_1\cos(\phi_1(t))$, then, $\dot{x}(t)=-A_1\sin(\phi_1(t)) \dot{\phi}_1 (t)$, which implies
\begin{equation}\label{prueba2}
A_1\sin(\phi_1(t))=-\frac{\dot{x}(t)}{\dot{\phi}_1(t)},
\end{equation}

where $\dot{\phi}_1(t)>0$ as we are assuming $\omega_1>0$.

Now, the ODE  in Theorem \ref{th:ode}(a) is easily obtained from (\ref{prueba2}), as follows,
$$\ddot{x}(t)=-A_1\cos(\phi_1(t))\dot{\phi}_1(t)-A_1\sin(\phi_1(t))\ddot{\phi_1}(t)=-x(t)\dot{\phi_1}(t)+\dot{x}(t)\frac{\ddot{\phi}_1(t)}{\dot{\phi}_1(t)}$$

Finally, let be $x(t)=\tan(\frac{\Phi(t)}{2})$, now using (\ref{prueba1}) it easily to show that
$$\dot{x}(t)=\frac{\omega_1}{2\cos^2(\frac{t-\alpha_1}{2}) }=\frac{\omega_1}{2}[1+\tan^2(\frac{t-\alpha_1}{2})]=\frac{\omega_1}{2}+\frac{1}{2\omega_1}x^2(t)$$
and Theorem \ref{th:ode} (b) follows. 

\bibliographystyle{apalike}

\bibliography{referenciasTotal}
\end{document}